\documentclass[epj,final]{svjour}

\usepackage{latexsym}
\usepackage{url}
\usepackage{amsfonts}
\usepackage{amsmath, amssymb}
\RequirePackage{graphicx}

\begin{document}
\title{On the Lambda-evolution of galaxy clusters}
\author{V.G.~Gurzadyan\inst{1,2}, A.A.~Kocharyan\inst{3},
A.~Stepanian\inst{1}
}                     
%
%
\institute{Center for Cosmology and Astrophysics, Alikhanian National Laboratory and Yerevan State University, Yerevan, Armenia \and
SIA, Sapienza Universita di Roma, Rome, Italy \and 
School of Physics and Astronomy, Monash University, Clayton, Australia}
\date{Received: date / Revised version: date}
%

\abstract{
The evolution of galaxy clusters can be affected by the repulsion described by the cosmological constant. This conclusion is reached within the modified weak-field General Relativity approach where the cosmological constant $\Lambda$ enables to describe the common nature of the dark matter and the dark energy. Geometrical methods of theory of dynamical systems and the Ricci curvature criterion are used to reveal the difference in the instability properties of galaxy clusters which determine their evolutionary paths. Namely, it is shown that the clusters determined by the gravity with $\Lambda$-repulsion tend to become even more unstable than those powered only by Newtonian gravity, the effect to be felt at cosmological time scales.     
} 
%
\maketitle

\section{Introduction}

The nature of the dark sector - dark matter and dark energy - remains a major puzzle for fundamental physics in spite of intense observational, experimental and theoretical investigations of the last decades. The recently sharpened $H$-tension, i.e. the discrepancy between the Hubble constant determinations from {\it Planck's} data and observations at lower redshift  \cite{Riess,Riess1}, activated the discussions regarding beyond $\Lambda$CDM and new physics \cite{Ver,DiV}. 

The modified weak-field General Relativity (GR) provides one of recent approaches to describing the dark sector \cite{G,GS1,GS2}. That modification is based on Newton's theorem on the identity of gravity of a sphere and of a point mass situated in its center and enables to conclude on the common nature of the dark matter and dark energy, both being described by the cosmological constant  \cite{GS1,GS2}. That approach also offers a solution to the $H$-tension \cite{GS3}. The cosmological constant $\Lambda$ within that approach acts as a  fundamental constant along with the gravitational constant $G$ \cite{GS4}, with consequences also for the Conformal Cyclic cosmology \cite{GP}. The non-particle nature of the dark matter is concluded in \cite{Popl}.

If the gravitational interaction at the galaxy cluster scales is defined not only by an attracting force but also by a repulsive force due to the cosmological constant, then the latter can influence the evolution of galaxy clusters. Analogous modified gravity effects are among the discussed ones with respect to various astrophysical systems, from celestial mechanical scales \cite{KEK} to those of cosmological structure formation \cite{E}, whereas the dedicated experimental tests of General Relativity (e.g. \cite{Tur,Ciu}) are still not reachable to those effects. Other testing opportunities are provided by the lensing of galaxies \cite{Col,GS5}.   

Here we will try to reveal a possible difference in the dynamical evolution of two types of galaxy clusters, those determined by usual Newtonian gravity (standard N-body problem) and those by modified gravity with the repulsion term.  We use geometrical methods of the theory of dynamical systems \cite{An,Arn} first applied to gravitational N-body dynamics in \cite{GS86} to describe their chaos and relaxation; for further application of those methods in General Relativity see \cite{K,AGK}.  The Ricci curvature criterion of relative instability, that we use here, was introduced in \cite{GK87} and has been applied to different types of gravitational systems (e.g. \cite{EZ}). Our aim is not the study of entire evolution of galaxy clusters affected by modified gravity, which will need extensive strategy of N-body simulations but to verify if the $\Lambda$ term is able to influence the cluster dynamics and the evolution. Such an approach appears informative for nonlinear systems, as known since the renown Fermi-Pasta-Ulam study \cite{FPU}.   

The results of our analysis indicate that the $\Lambda$-gravity does affect the instability features of galaxy clusters.  We note that previously it was shown that the cosmological constant is able to introduce a time arrow for the system \cite{AG}.

\section{Newton's theorem and $\Lambda$}

Proceeding from the Newton's theorem on the ``sphere-point" identity and the resulting weak-field modification of General Relativity, one arrives to the metric \cite{GS1} \textbf{$(c=1)$}
\begin {equation} \label {mod}
g_{00} = 1 - \frac{2 G m}{r} - \frac{\Lambda r^2}{3}; 
\qquad g_{rr} = \left(1 - \frac{2 G m}{r} - \frac {\Lambda r^2}{3}\right)^{-1}.
\end {equation} 

The general functions  for force and potential i.e. $U(r)$ and $\mathbf{F}(r)$ satisfying Newton's theorem used for the above weak-field limit of GR have the form (for derivation and discussion see \cite{G1,G,GS1}) 
\begin{equation}
\label{FandU}
U(r) = -\frac{A}{r} - \frac{B}{2}r^2;
\qquad 
\mathbf{F}(r) = -\nabla_rU(r) = \left(-\frac{A}{r^2} + Br\right)\hat{\mathbf{r}}\ .
\end{equation}   
Here the second term leads to the cosmological term in the solutions of Einstein equations, so that the cosmological constant $\Lambda$ enters also the weak-field GR regimes, e.g. in the Hamiltonian dynamics of galaxy clusters \cite{GS3}. 

Crucial feature of the force law of Eq.(\ref{FandU}) is that it defines non-force-free field inside a spherical shell, contrary to Newton's gravity law when the shell has no influence in its interior. In this regard we mention the observational indications that the galactic halos do determine the properties of galactic disks \cite{Kr}. The weak-field GR thus can be used to describe the observational features of galactic halos \cite{G,Ge}, of groups and clusters of galaxies \cite{GS2}.

\section{Ricci curvature}

The Lagrangian for N-body system interacting by the $\Lambda$-potential (from Eq.(\ref{FandU})) is
\begin{eqnarray}
\label{Lag}
  L(r,v) &=& \tfrac{1}{2}\sum_{a=1}^{N}m_av^2_{a} - U(r),\\
  U(r)   &=& -\sum_{a=1}^{N}\sum_{b=1}^{a-1}\frac{Gm_am_b}{|r_a-r_b|}
             -\frac{\Lambda}{6}\sum_{a=1}^Nm_a|r_a|^2.
\end{eqnarray}

According to the criterion of relative instability defined in \cite{GK87}\footnote{VG and AK recall with gratitude the discussion of this criterion with Vladimir Arnold and his important comments, as mentioned in \cite{GK87}.}, among two systems the more unstable is the one with smaller negative Ricci curvature
\begin{equation}
\label{r}
\mathfrak{r} = \tfrac{1}{3N}\inf_{0\leq s\leq s_*}\mathfrak{r}_u(s),
\qquad \mathfrak{r} < 0,
\end{equation}
within $0\leq s\leq s_*$ interval of geodesic in the configurational space.

This criterion follows from the equation of geodesic deviation (Jacobi-Levi-Civita) equation \cite{An,Arn} averaged via the
deviation vector
\begin{equation}
\frac{d^2z}{ds^2}= - \frac{1}{3N}\mathfrak{r}_u(s)+\langle\parallel\nabla_un\parallel^2\rangle,
\end{equation}
where
$$
n=z\hat{n}, \qquad \parallel \hat{n}\parallel^2=1,
$$
and $\mathfrak{r}_u(s)$ is the Ricci  curvature  in  the  direction  of  the velocity of the geodesic $u$, and
\begin{equation}
\mathfrak{r}_u(s)= \frac{\text{Ric}(u,u)}{u^2}
=\sum_{\mu=1}^{3N-1}
K_{\mathbf{e}_{\mu},u}(s),\qquad 
(\mathbf{e}_{\mu}\bot u,\ \mathbf{e}_{\mu}\bot\mathbf{e}_{\nu},\ \mu\ne\nu).
\end{equation}
The Ricci tensor for N-body system yields \cite{GS86,GK87}
\begin{eqnarray}
\text{Ric}_{\alpha\beta} &=& -\frac{1}{2}\frac{\Delta W}{W} g_{\alpha\beta}
-\frac{(3N-2)}{2}\frac{W_{\alpha\beta}}{W}
+ \frac{3(3N-2)}{4}\frac{W_{\alpha}W_{\beta}}{W^2}
-\frac{(3N-4)}{4}\frac{\|dW\|^2}{W^2}g_{\alpha\beta},
\end{eqnarray}
where $g_{\alpha\beta}=m_a\delta_{\alpha\beta}$, and 
\begin{equation}
W=E-U=E +\sum_{a=1}^{N}\sum_{b=1}^{a-1}\frac{Gm_am_b}{|r_a-r_b|}
             +\frac{\Lambda}{6}\sum_{a=1}^Nm_a|r_a|^2 =\tfrac{1}{2}\sum_{a=1}^Nm_a v_a^2.
\end{equation} 

For the Lagrangian of Eq.(\ref{Lag}) the latter is 
\begin{align}
\label{Ric}
{\text{Ric}}(v,v)
    &=\frac{(3N-2)}{2W}\ \sum_{c=1}^{N}\sum_{^{a=1}_{a\ne c}}^N
    \frac{Gm_cm_a}{\rho_{ca}^3}\left(v_c\cdot v_{ca}
        -3\frac{(r_{ca}\cdot v_c) (r_{ca}\cdot v_{ca})}{\rho_{ca}^2}\right)\nonumber\\
    &+\frac{3(3N-2)}{4W^2}\left(- \sum_{c=1}^{N}\sum_{^{a=1}_{a\ne c}}^N 
    Gm_cm_a \frac{r_{ca}\cdot v_c}{\rho_{ca}^3} 
    +\frac{\Lambda}{3}\sum_{c=1}^N m_c r_c\cdot v_c\right)^2\nonumber\\
    &-\frac{(3N-4)}{2W}\sum_{c=1}^N m_c
    \left|-\sum_{{}^{a=1}_{a\ne c}}^NGm_a \frac{r_{ca}}{\rho_{ca}^3}
    +\frac{\Lambda}{3}r_c\right|^2 -\frac{2(3N-1)}{3}\Lambda\ ,
\end{align}
where
$r_{ab}=r_a-r_b$,
$v_a=\dot{r}_a$,
$v_{ab}=v_a-v_b$, 
$\rho_a=|r_a|$,
$\rho_{ab}=|r_{ab}|$, for any two vectors $\mathbf{e}_1$ and $\mathbf{e}_2$ we have 
$\mathbf{e}_1\cdot\mathbf{e}_2=\delta_{ij}e_1^ie_2^j$.\newline

\section{Results}

We simulated the dynamics of two types of N-body spherical systems of typical galaxy cluster parameters, one defined by Newtonian gravity, the other defined by an additional $\Lambda$-potential i.e. by Lagrangian Eq.(\ref{Lag}). The Ricci curvature was estimated for both, to see if the instability properties of the both systems according to the criterion Eq.(\ref{r}) do reveal differences during their evolution over cosmological time scale.

To simulate systems with typical parameters of galaxy clusters we used a spherical distribution of $N=1,000$ particles (galaxies), each of mass $m=10^{11} M_\odot$, inside a sphere of $R=1.5$ Mpc. The velocities were defined by considering such galaxy clusters (i.e. of $1,000$ members)  as semi-virialized configurations, i.e. 
\begin{equation}\label{vir}
\sigma^2 = \frac{GNm}{R} = \frac{GM}{R},
\end{equation} 
where $\sigma^2$ is the velocity dispersion of galaxies of the cluster and $M$ is the total mass of the cluster. Consequently, the dynamical time scale for a typical cluster in Newtonian and $\Lambda$-modified regime will be
\begin{equation}\label{tdyn}
t_G = \left(\frac{2R^3}{GM}\right)^{1/2} = 3.86 \, \text{Gyr}, 
\quad t_{G\Lambda} = \left(\frac{2 R }{\frac{GM}{R^2} - \frac{\Lambda R}{3}}\right)^{1/2} = 3.90 \, \text{Gyr}\, .
\end{equation}
The results of computations using Eqs.(\ref{r})-(\ref{Ric}) are shown in Fig.1. Comparing both Newtonian and $\Lambda$-modified gravity it turns out that the behavior of Ricci curvature $\mathfrak{r}$ for both cases are similar to each other (Fig.1). However, while the values of Ricci curvatures practically coincide at time scales up to around 2 Gyrs, at later phases those values for pure Newtonian case are systematically larger than for $\Lambda$-modified gravity. According to the criterion Eq.(\ref{r}) in view of the fact that the value  $\mathfrak{r}_{G\Lambda}$ is smaller than  $\mathfrak{r}_{G}$ - both having negative $\it infimum$ according to criterion Eq.(\ref{r}) within the cosmological time interval 
[2.4 Gyr - 17.4 Gyr] - one can conclude that spherical Newtonian systems which at large $N$ limit are known to be exponentially unstable (chaotic) \cite{GS86}, become even more unstable with the $\Lambda$-term in the gravity force Eq.(\ref{FandU}).  

Then we performed the same analysis for systems of parameters of superclusters, using the data of the Virgo Supercluster. Note, that there is a principal difference between this case and those of galaxy clusters for $\Lambda$-modified gravity. Namely, from Eq.(\ref{mod}) and Eq.(\ref{FandU}) one can define a critical distance scale for a system, where the repulsive term of $\Lambda$ becomes dominant \cite{GS3} over the Newtonian gravity 
\begin{equation}\label{crit}
r_{crit}^3 = \frac{3GM}{\Lambda}\, .
\end{equation}

For structures of smaller than superclusters' scale this radius lies outside the configuration which means that the role of the $\Lambda$ term in its properties is suppressed.  But for superclusters of scales larger than $r_{crit}$  the role of $\Lambda$-term can be felt in the dynamics of galaxies \cite{GS3}. For Virgo Supercluster that critical radius yields around 12.66 Mpc. In this regard, we checked the behavior of Ricci curvature for three different cases. First, we analysed a system of parameters of the Virgo Supercluster, i.e. $R$= 16.5 Mpc, $N=1,480$, and $M=1.48$ $\times$ 10${}^{15}$ $M_\odot$. The results i.e. the difference of Ricci curvature for Newtonian and $\Lambda$-modified gravity, are given in Fig.3. Then we studied two different cases, i.e.  with the same mass and number of particles but for different radii i.e. $R=18$ ($>$12.66) Mpc and $R=10$ Mpc ($<$12.66). For latter two cases the results are shown in Figs.4-5. It is interesting that for both, 18 Mpc and 16.5 Mpc (both exceeding the critical distance 12.66 Mpc), as time goes on the difference of Ricci is increasing and even for $R=18$ Mpc, it becomes positive, which can be interpreted as tending to free particle system. While for the bound structure $R=10$ Mpc the Ricci curve shows a tendency to decrease, i.e. indicating the unstable N-body system.

\begin{figure}[!htbp]
  \centering
  \includegraphics[width=100mm]{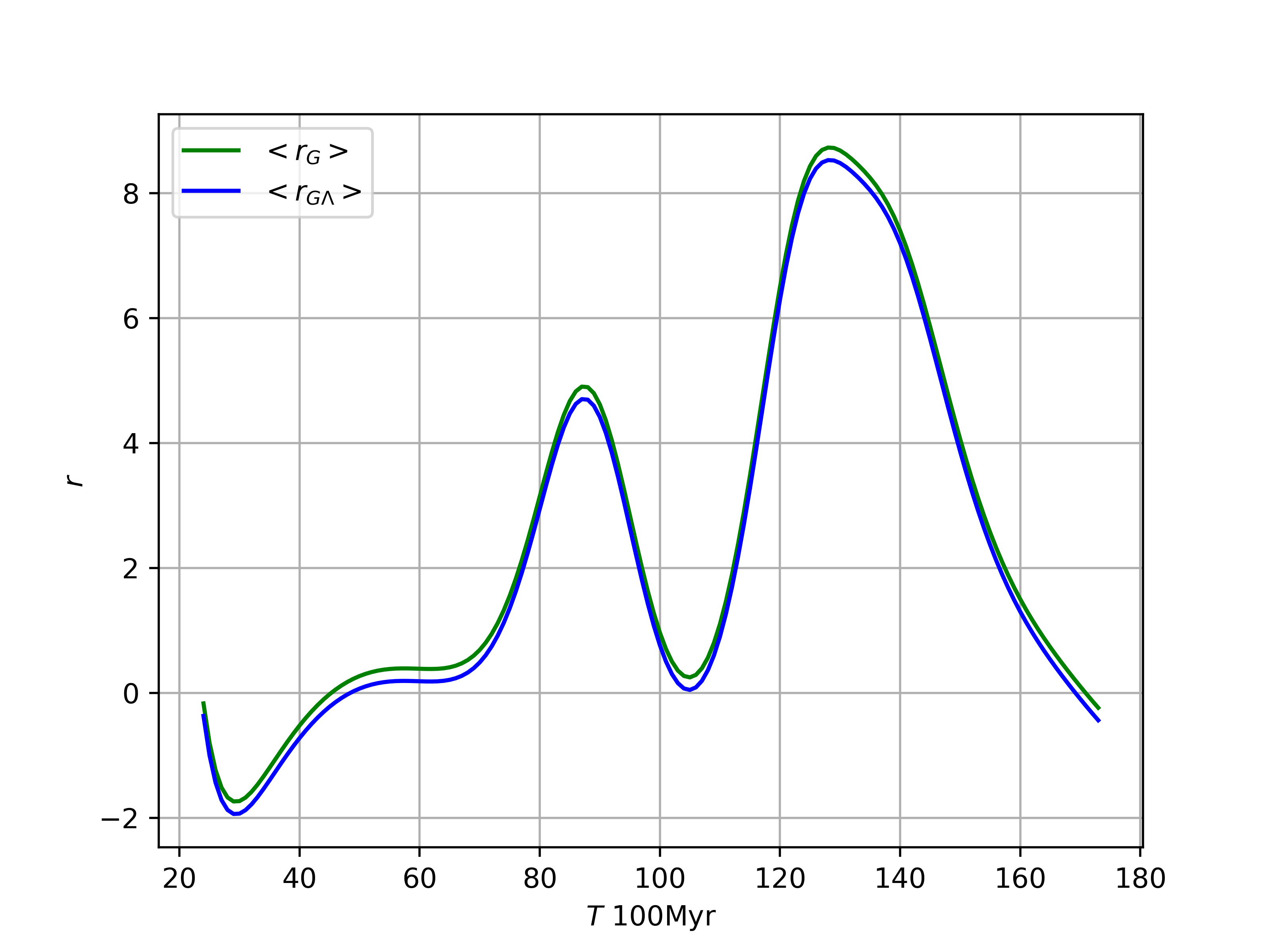}
  \caption{The Ricci curvature variation vs cosmological time for galaxy cluster parameters in Newtonian (green) and $\Lambda$-modified gravity (blue) regimes.}
\end{figure}

\begin{figure}[!htbp]
  \centering
  \includegraphics[width=100mm]{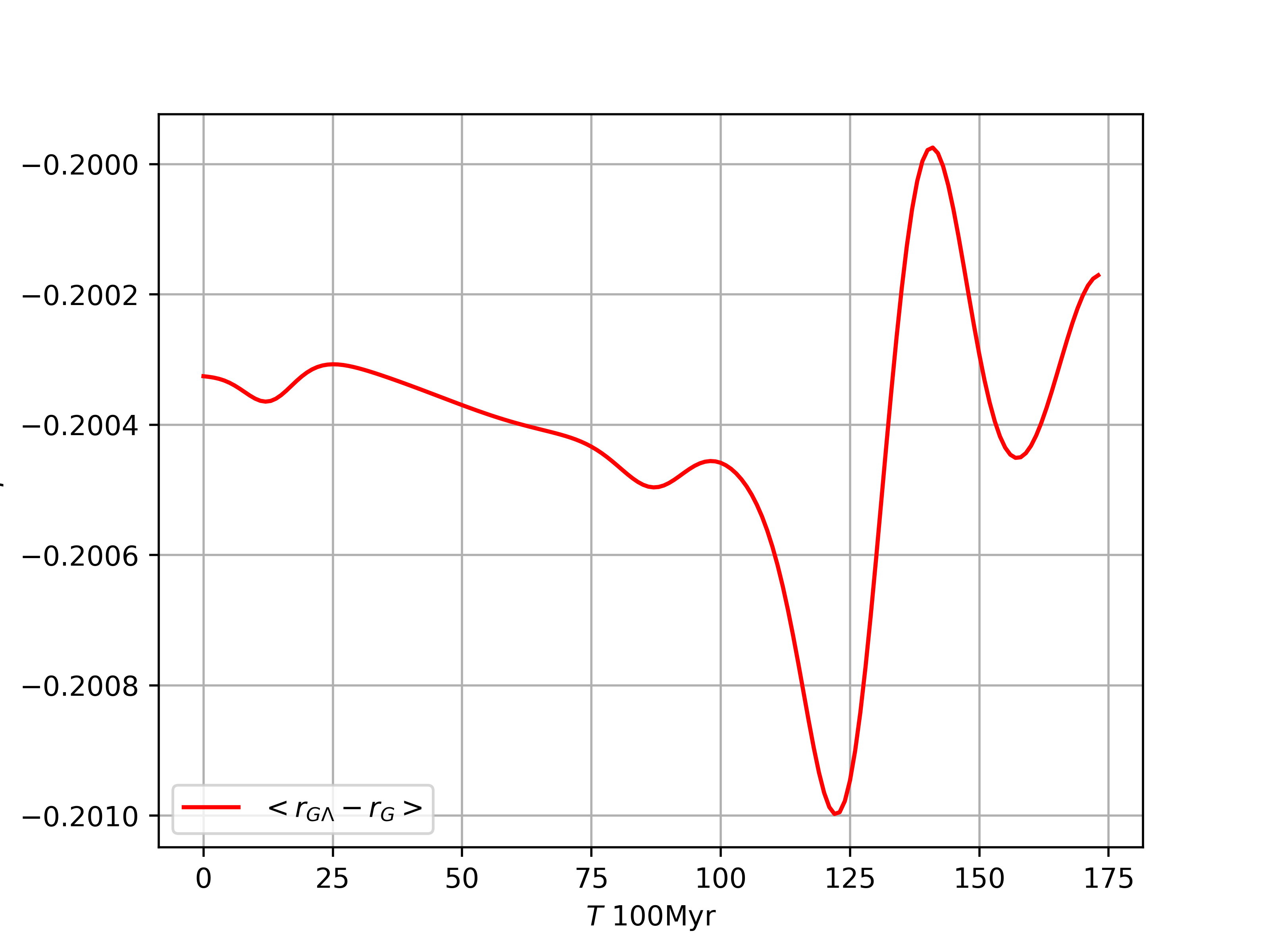}
  \caption{The difference of Ricci curvature values of curves in Fig.1, i.e. at Newtonian and $\Lambda$-modified gravity laws.}
\end{figure}

\begin{figure}[!htbp]
  \centering
  \includegraphics[width=100mm]{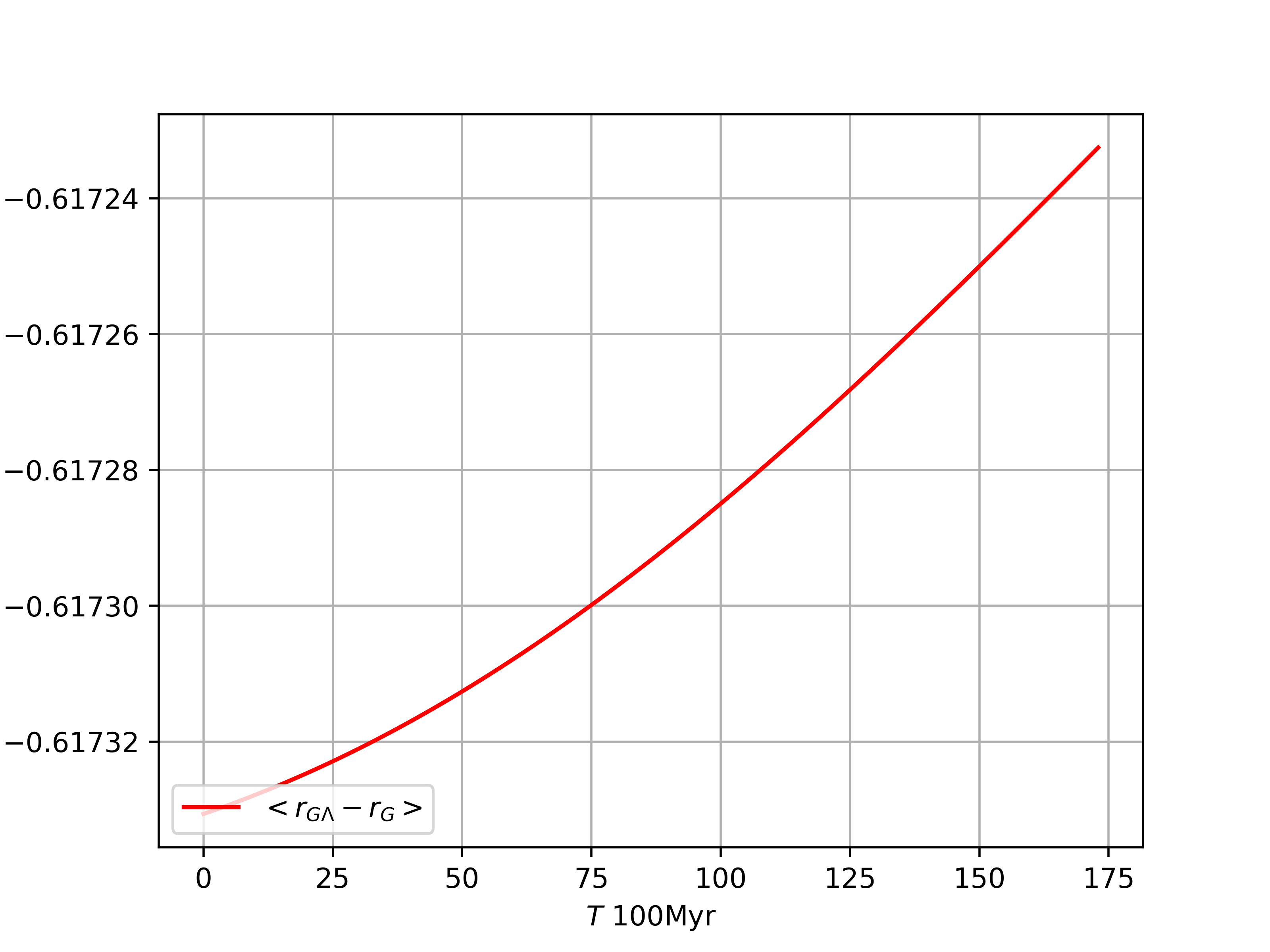}
  \caption{The same as in Fig.2 but for supercluster (Virgo) parameters.}
\end{figure}

\begin{figure}[!htbp]
  \centering
  \includegraphics[width=100mm]{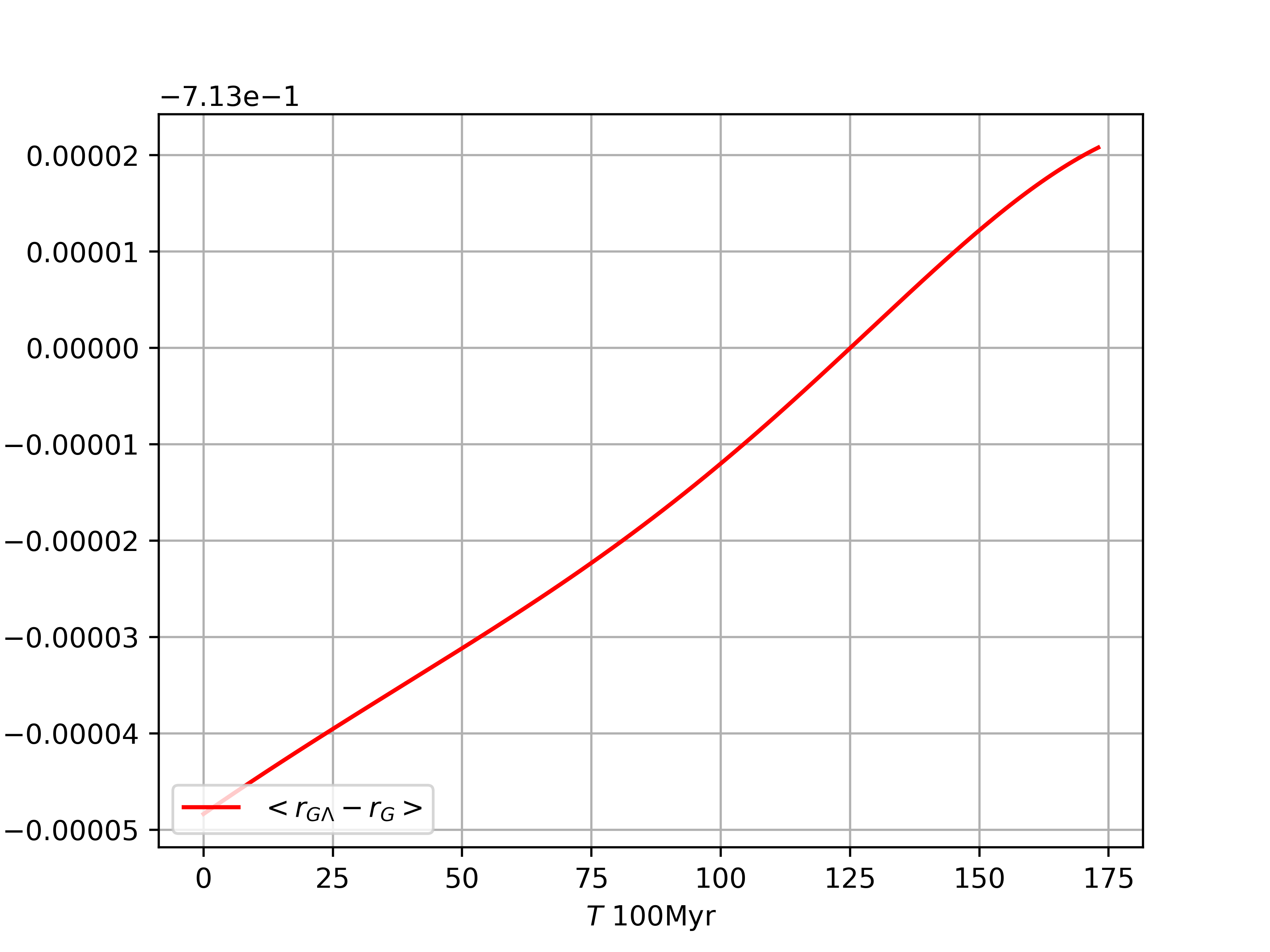}
  \caption{The same as in Fig.3 but for $R=18$ Mpc.}
\end{figure}

\begin{figure}[!htbp]
  \centering
  \includegraphics[width=100mm]{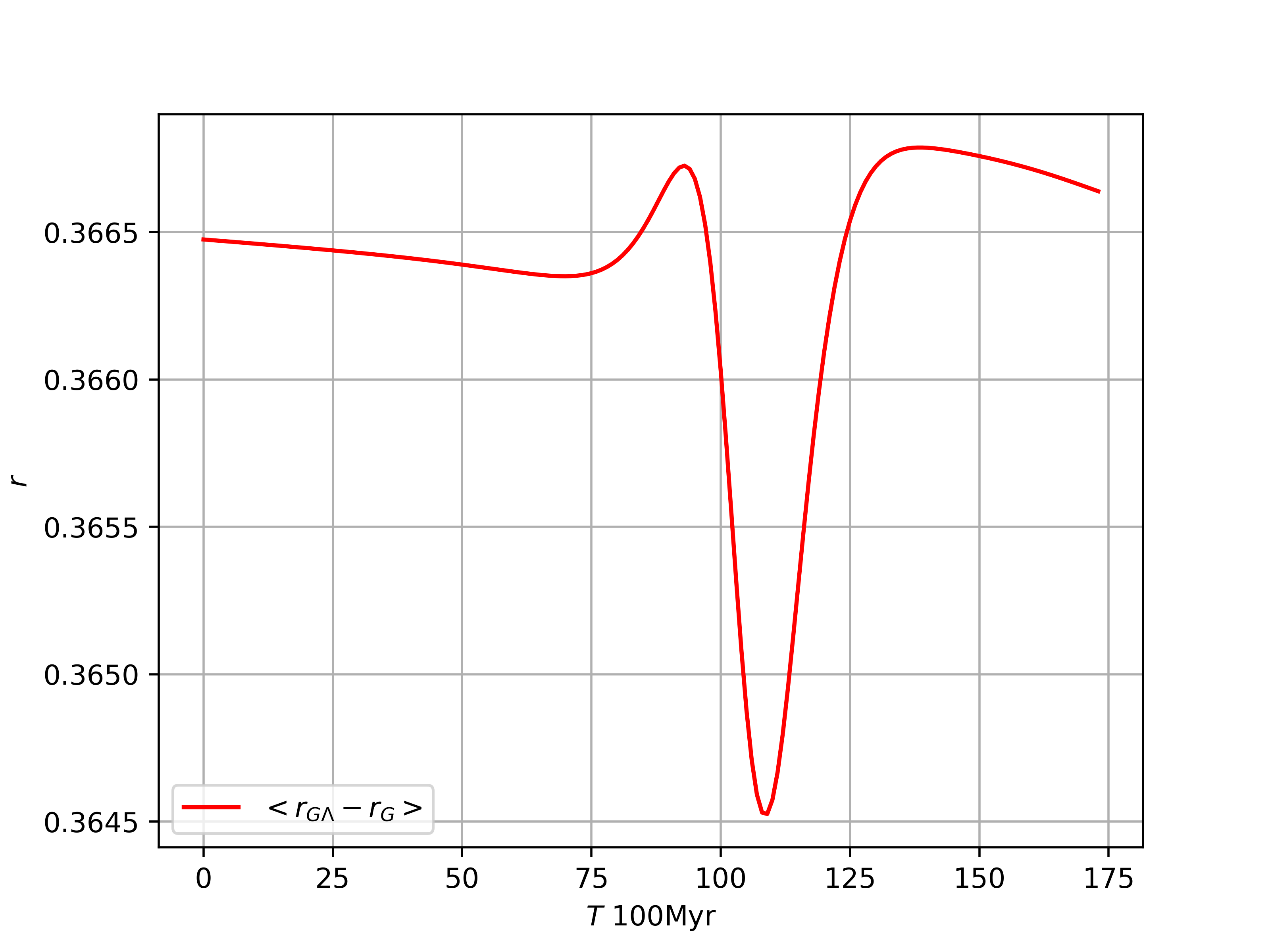}
  \caption{The same as in Fig.3 but for $R=10$ Mpc.}
\end{figure}

\section{Conclusions}

The weak-field GR modified based on the Newton theorem enabled one the common description of the dark matter and dark energy \cite{G,GS1}, as weak-field and cosmological manifestation of GR, respectively, both determined by the cosmological constant. 

As further step to probe that GR modification, we analysed its possible role in the evolution of the galaxy clusters, i.e. at spatial scales where the repulsive gravity term becomes non-negligible. We used the Ricci curvature criterion to follow the comparative instability of two type of spherical systems, i.e. those evolving according to modified $\Lambda$-gravity of Eqs.(\ref{mod}) and (\ref{FandU}) with respect to usual Newtonian systems.

Our main conclusions can be formulated as follows:

(a) the  studied types of systems of galaxy cluster parameters do reveal discrepancy in their instability properties during the evolution, namely, the $\Lambda$-modified gravity systems tend to become more unstable with respect to those described by Newtonian law. The discrepancy starts to be visible at cosmological times i.e. at time scales exceeding roughly  2 Gyr.

(b) the supercluster (Virgo) parameter systems reveal differences in their instability properties depending on their spatial scales. Namely, at distance scales where the $\Lambda$-term dominates over the Newtonian gravity, the systems tend to free particle systems at cosmological time scales, while at smaller distances their behavior remains unstable as of the galaxy clusters, as expected.  

We note that $f(R)$-gravity, "beyond Horndeski" covariant Galileon models have been already used to describe the observable features of clusters of galaxies \cite{Cap1,Cap2,Cap3}, thus indicating the suitability of the latter for testing of modified gravity theories.    

The study of the evolutionary effects of galaxy clusters  at dedicated numerical simulations (including using advanced methods of the theory of dynamical systems \cite{AGK1}) can provide additional tests to $\Lambda$-gravity as the weak-field limit for General Relativity.

\section{Acknowledgment}
AS is partially supported by the ICTP through AF-04.

\end{document}